\documentclass[prl,amsmath,amssymb,floatfix]{revtex4}
\usepackage[latin9]{inputenc}
\usepackage{color}
\usepackage[caption=false]{subfig}
\usepackage{verbatim}
\usepackage{graphicx}
\usepackage{esint}
\usepackage{epstopdf}

\makeatletter
\@ifundefined{textcolor}{}
{%
 \definecolor{BLACK}{gray}{0}
 \definecolor{WHITe}{gray}{1}
 \definecolor{ReD}{rgb}{1,0,0}
 \definecolor{GReeN}{rgb}{0,1,0}
 \definecolor{BLUe}{rgb}{0,0,1}
 \definecolor{CYAN}{cmyk}{1,0,0,0}
 \definecolor{MAGeNTA}{cmyk}{0,1,0,0}
 \definecolor{YeLLOW}{cmyk}{0,0,1,0}
}

\usepackage{epsfig}
\usepackage{dcolumn}
\usepackage{bm}
\def\(({\left(}
\def\)){\right)}
\def\[[{\left[}
\def\]]{\right]}

\newcommand{\be}{\begin{equation}}
\newcommand{\ee}{\end{equation}}
\newcommand{\bea}{\begin{eqnarray}}
\newcommand{\eea}{\end{eqnarray}}

\makeatother

\begin{document}

\title{Quantum bound to chaos and the semiclassical limit}

\author{ Jorge Kurchan$^{1}$}

\affiliation{
$^{1}$Laboratoire de Physique Statistique, \'Ecole Normale Sup\'erieure, PSL Research University; 
Universit\'e Paris Diderot Sorbonne Paris-Cit\'e; Sorbonne Universit\'es UPMC Univ Paris 06; CNRS; 24 rue Lhomond, 75005 Paris, France.}

\begin{abstract}
We discuss the quantum bound on chaos in the context of the free propagation of a particle in an arbitrarily curved surface at low temperatures.
The  semiclassical calculation of the Lyapunov exponent can be performed in much the same way as the corresponding one for the `Loschmidt echo'.
The bound   appears here as the impossibility to scatter a wave, by effect of the curvature,   over characteristic lengths smaller than the deBroglie wavelength.   
\end{abstract}
\maketitle
\vspace{0.5cm}

In a recent paper a bound to chaos, imposed by quantum dynamics, was derived. Although the original context of the discussion was that 
of Black Holes and String Theory, the result itself is a general elementary property of quantum
mechanics, and the proof only relies on Shr\"odinger's equation plus reasonable `clustering' properties of operators \cite{mss} (see also \cite{Hayden,Susskind}).
In this note, I  discuss the appearance of this result in a setting in which  it is an   intuitive property of a wave dynamics in
a  scattering medium.  

\section{ Fidelity}

Consider a system evolving with a Hamiltonian $H$ for a time $t$, and then backwards with a slightly different hamiltonian $-H+\delta H$: a `Loschmidt echo' \cite{Peres,Pastawski,Prosen} experiment (see, in the present context, \cite{Campisi,Miyaji}). A particular case is when at time $t$, the system is subjected to a rapid `kick' with Hamiltonian $\delta H(t')=B \delta(t'-t)$,
and subsequently  is evolved back for a time $t$ with $-H$. 
In classical mechanics, the returning trajectory is the time-reversed of the initial trajectory, except that its beginning has been  perturbed. It is clear then
that if the system is chaotic, and the time $t$ is long enough, the return point differs from the initial point by a magnitude $C e^{\lambda t}$, where $\lambda$ is the Lyapunov exponent that does not depend on the nature of the kick (for $t\lambda \gg 1$), and $C$ is a constant that does.
 
 An analogous result holds for a quantum system if one prepares it in a localized state,  and evolves it with a Hamiltonian and back with a
 slightly perturbed Hamiltonian.  If one measures appropriately the overlap between the initial and the returning wavefunction, there is a regime where  
  this difference scales exponentially in time, with the (quantum Lyapunov) exponent independent of the perturbation between the outgoing and ingoing Hamiltonians \cite{Levstein}.
Several years ago, Jalabert and Pastawski \cite{Pastawski} showed that for a semiclassical system, and for a given range of perturbation and times, this quantum exponent corresponds
to the classical one (although  an `annealed' version thereof, see discussion in \cite{Beenakker}).

Consider a Hermitean operator $A$, which we shall assume  is concentrated in some region of phase space, as seen, for example, in its coherent-state or Wigner representation.
We introduce  a measure of the `fidelity' $F= \mbox{Tr} [A_{tran}A]$, comparing  the operator $A$ with the   one $A_{tran}$ transformed  by Loschmidt echo trajectory. For a `kick' perturbation, we have:
\begin{equation}
F= \mbox{Tr} [A_{tran}A]=\mbox{Tr} \left\{\left[ e^{i\frac{t}{\hbar}H}  e^{i \frac{\delta}{\hbar} B } e^{-i\frac{t}{\hbar}H} \right] A \left[e^{i\frac{t}{\hbar}H}  e^{i \frac{\delta}{\hbar} B } e^{-i\frac{t}{\hbar}H} \right]^\dag A 
\right\} = 
\mbox{Tr} \left\{ A e^{i\frac{t}{\hbar}H}  e^{i \frac{\delta}{\hbar} B } e^{-i\frac{t}{\hbar}H}  A e^{i\frac{t}{\hbar}H}  e^{-i \frac{\delta}{\hbar} B } e^{-i\frac{t}{\hbar}H}  \right\}
\end{equation}

If $\delta=0$ we clearly get $F=\mbox{Tr} [A^2]$. Putting 
\begin{equation}
B(t)=e^{i\frac{t}{\hbar}H}  B  e^{-i\frac{t}{\hbar}H} 
\end{equation}
and developing in powers of $\delta$ (linear response)  $ e^{i \frac{\delta}{\hbar} B(t) } =1+i \frac{\delta}{\hbar} B (t)-\frac12
\frac{\delta^2}{\hbar} B^2(t)+...$ we get:
\begin{equation}
 \mbox{Tr} [A^2] - \mbox{Tr} [A_{tran}A] =  - \frac{\delta^2}{\hbar^2} \mbox{Tr} \left\{ B(t) A B(t) A  \right\} +  \frac{\delta^2}{\hbar^2} \mbox{Tr} \left\{ B^2(t) A^2  \right\}=   -\frac{\delta^2}{2\hbar^2} \mbox{Tr} \left\{ [B(t),A]^2 \right\}
\end{equation}
The measure we are seeking is thus \cite{mss}
\begin{equation}
F_1= - \mbox{Tr} \left\{ [B(t),A]^2 \right\} \label{crochet}
\end{equation}
which is positive since it is minus the trace of an anti-Hermitean operator squared.
We need the operator $A$ to be concentrated, in order to localize our measure in a region of phase space, and to make the trace converge. Two choices have been proposed. A traditional one
is is to consider a localized wavefunction $|\psi\rangle$ \cite{Pastawski,Prosen} and:
\begin{equation}
A=|\psi\rangle \langle \psi| \hspace{1cm} \Rightarrow \hspace{1cm} F= |\langle \psi |e^{i\frac{t}{\hbar}H}  e^{i \frac{\delta}{\hbar} B } e^{-i\frac{t}{\hbar}H} |\psi \rangle|^2
\end{equation}
(here in the `kick' version,  although more commonly with a continuous perturbation $\delta H$).
A more recent one \cite{mss}, that lends itself better for the study of an extensive system at temperature $T=\frac 1\beta$, is obtained by choosing a smooth (possibly unlocalized) Hermitean operator $W$ and to render it concentrated on an energy shell, at least in the thermodynamic
limit,  by putting in (\ref{crochet}): 
\begin{equation}
A \propto e^{-\frac{\beta}{4} H } W e^{-\frac{\beta}{4} H } \hspace{.3cm} \Rightarrow \hspace{.3cm} F_1\propto  - \mbox{Tr}\left\{
  B(t)  e^{-\frac{\beta}{4} H } W e^{-\frac{\beta}{4} H }B(t) e^{-\frac{\beta}{4} H } W  e^{-\frac{\beta}{4} H }  \right\}
   +\mbox{Tr}\left\{
  B^2(t)  e^{-\frac{\beta}{4} H } W e^{-\frac{\beta}{2} H } W e^{-\frac{\beta}{4} H }
 \right\}
 \label{lyap2}
\end{equation}
The symbol $\propto$ means that $A$ is normalized by $\mbox{Tr } e^{-\beta H/2}$. 
The second term only contributes a two-time function and  its precise form is not relevant beyond canceling constants, we keep it for convenience here.
(The regularization strategy of splitting the equilibrium density operator in factors -- here four of them -- is standard in Chemical Physics, see \cite{Miller}.)
For this form, in some cases there is a Lyapunov regime:
\begin{equation}
F_1(t) \sim f_1 e^{\lambda t}
\end{equation}
 In this context, it has  been shown that 
there is a  bound to chaos \cite{mss} of purely quantum nature:
\begin{equation}
\beta \hbar \lambda \leq 2 \pi
\label{bound}
\end{equation}

\section{Model}

In the semiclassical case, where there is no apparent limitation, 
the question immediately arises of how does quantum mechanics intervene \footnote{This was discussed, for the case of billiards  in \cite{mss}, and also in \cite{Garcia} in relation to the entanglement rate. }.
We start by considering a Hamiltonian system of the form 
\begin{equation}
H_o= \sum_1^{N'}p_i^2/2m +V(q) \label{cosa}
\end{equation}
 in the semiclassical limit $\hbar \rightarrow 0$.
For the bound to be effective, we need the product $\beta \hbar$ to be finite,
and this may only be achieved in the semiclassical limit for very low temperatures, corresponding to the lowest classical energies of the system.
It is clear that if the classical ground state does not have dimension larger than two, there will be no chaos: the system will sit on
  the classical ground state and perform vibrations and quantum fluctuations around it.
In order to render the system interesting from our point of view, we consider a potential such that the minimum of $V(q)$ is a manifold of dimension $N<N'$.
Thus, even at the lowest energies, the system will have long-distance motion  `at the bottom of the well'.
Since we are interested in such a motion, we may concentrate on it by working with the quantum mechanics in the  curved
space of minima of the potential (a `quantum wire' \cite{wires} or `quantum conducting surface' problem). 
The Hamiltonian restricted to that manifold is constructed as follows: we set up a system of $N<N'$ curvilinear coordinates $q_i$ of the manifold of minima, and the associated momenta $p_i$, $i=1,...,N$  \cite{Ikegami}.
Let the metric tensor be $g_{ij}(q)$ so that length element $\delta \ell^2 = g_{ij} dq_i dq_j$. The Hamiltonian is proportional to the Laplace Beltrami operator \cite{Ikegami}
\begin{equation}
H= -\frac{\hbar^2}{2m} g^{-\frac12} \partial_i g^{ij} g^{\frac12} \partial_j=  \frac{\hbar^2}{2m}g^{-\frac14} Q_i^\dag g^{ij} Q_j g^{\frac14} 
\label{hamquan}
\end{equation}
where $g=\det\{g_{ij}\}(q)$ and $Q_i= g^{\frac14} \partial_i g^{-\frac14}$. The r.h.s. shows that the spectrum is positive semidefinite, including zero
if and only if the system is bounded. 
The imaginary-time evolution with Hamiltonian (\ref{hamquan}) is just diffusion on the constrained surface. The lowest eigenvalues
of  (\ref{hamquan}) indeed correspond to the inverse thermalization times associated with this diffusion process, the eigenvalue zero to the flat invariant measure.
Equation (\ref{lyap2}) can be viewed as an expectation of various observables (correlations and responses) of a process with four diffusion stretches  lasting $\beta/4$.

A two-dimensional version of the constant negative curvature surface has been extensively studied
as the simplest, and one of the few solvable chaotic systems, as  was discussed in detail by Balazs and Voros \cite{Voros}, 
\begin{equation}
H=  -\frac{\hbar^2}{8mR^2} ( 1- q_1^2-q_2^2)^2  \left[\frac{\partial^2}{\partial q_1^2} +  \frac{\partial^2}{\partial q_2^2}\right]
\label{hyper}
\end{equation}
where $R$ is the `radius' of the hyperboloid. The classical Lyapunov exponent is $\lambda=\sqrt{2E/mR^2}$. Expectation values may be computed  analytically,
because eigenbases are known. We shall
not do this interesting calculation here, but keep the example in mind throughout.

\vspace{.3cm}

{\bf Classical Limit}

\vspace{.3cm}

In the classical limit, the Hamiltonian is given by the kinetic energy of the constrained motion:
  \begin{equation}
H= \frac{1}{2m} p_i g^{ij}  p_j
\label{hamcla}
\end{equation}
and the Lagrangian:
 \begin{equation}
L(q,\dot q)= \frac{m}{2} \dot q_i g_{ij}   \dot q_j
\end{equation}
  where $g_{ij} $ is the inverse of $g^{ij}$. The geodesic lengths are then given by the extrema of:
  \begin{equation}
  \mbox{Length} = \int dt L^{1/2} 
  \end{equation}
  Because $L$ is a constant of motion, the Lagrange equations for geodesics and for the action lead to proportional equations. 
  We hence conclude that the classical trajectories are {\em geodesics} in the coordinate space of the $q_i$, traversed with a velocity 
  that is proportional to the square root of the (purely kinetic) energy. 
  
  If the system is bounded, we may compute the equilibrium quantities via:
  \begin{equation}
  Z= \int dq dp \; e^{-\beta H} \propto \left(mT \right)^{\frac N2}\int dq \;\; g^{\frac 12} \;\; \Rightarrow \;\;\;   \left\langle \frac EN \right\rangle_{class. \; equil.}  =\langle {\cal{E}} \rangle_{class. \; equil.}=\frac T2  
  \end{equation}
 $ {\cal{E}}$ is the energy per degree of freedom.

  It is interesting to observe that 
  in going from (\ref{cosa}) to its low-energy version (\ref{hamcla}), a kind of reparametrization invariance appears. Given two points in  {\em configuration} space and two times 
  $t_1$ and $t_2$, there is a trajectory in space that will join the points in time $t_1$ and another in $t_2$ just  by running over the same path at two different speeds.
  Note that this property was absent in the original system  (\ref{cosa}).
   Time-reparametrization plays an important role
  in quantum glass models that saturate the bound (\ref{bound}) \cite{syk,Kitaev,ms} 
  
  \begin{figure}
\centering \includegraphics[angle=0,width=5cm]{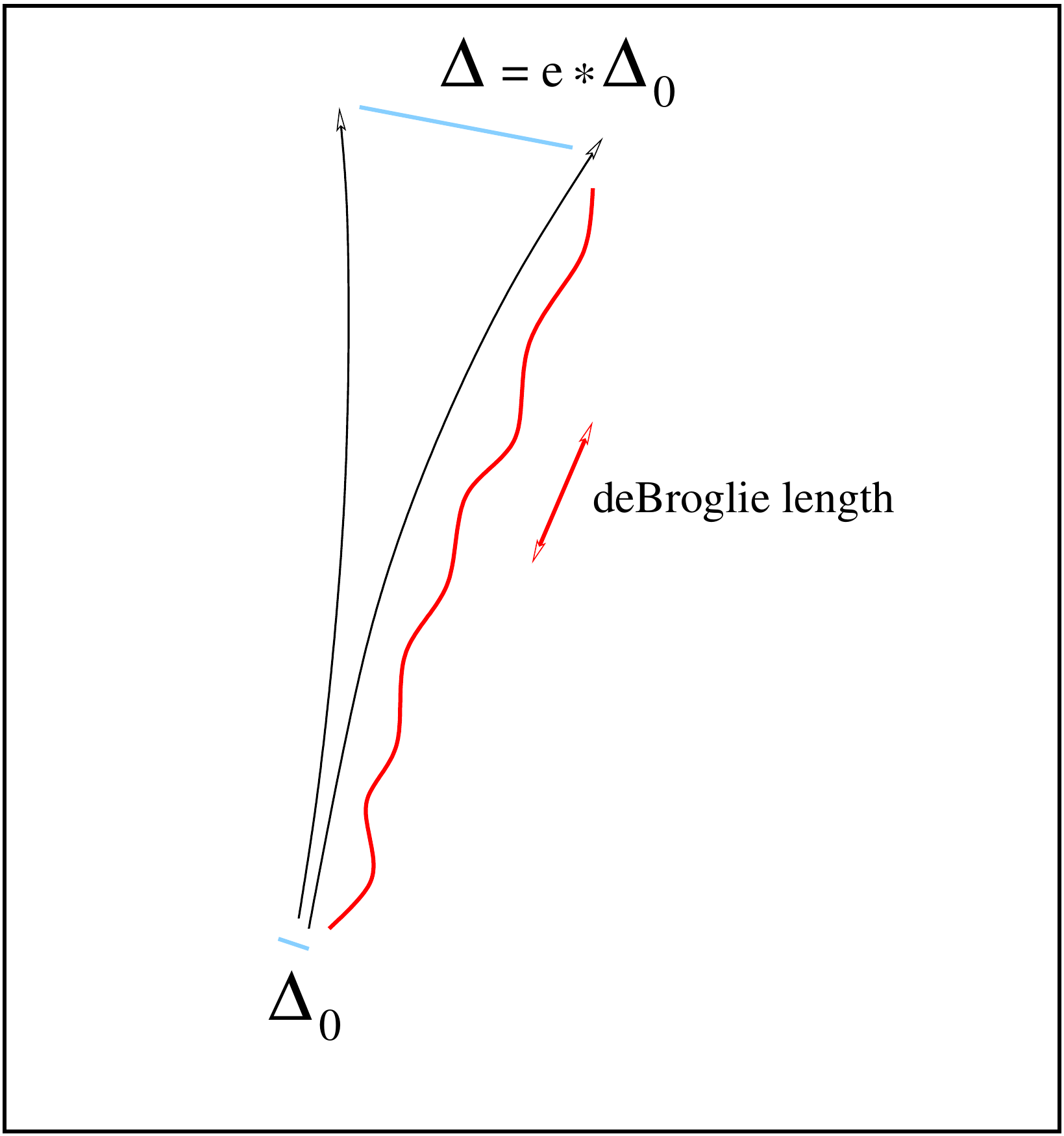} \protect\protect\protect\protect\caption{Geometric vs. deBroglie lengths.}
\label{remsketch} 
\end{figure}

  \vspace{.3cm}

{\bf Classical Lyapunov exponent and Geodesic Deviation}

\vspace{.3cm}
  
  Let us now assume that the ground state manifold is such that the geodesics are {\em chaotic}. As mentioned above, an  example is the constant negative curvature surface  in
two dimensions \cite{Voros}.
  Two nearby trajectories {\em in phase space} $(q_i,p_i)$ and $(q_i+\delta q_i,p_i+\delta p_i)$ will separate exponentially
  \begin{equation}
  \sqrt{ \sum_i [(\delta q_i)^2 + (\delta p_i)^2](t)}  \propto \;\; e^{\lambda t}
  \end{equation}
  It is easy to write a differential equation for the evolution of the phase-space Lyapunov vector $(\delta q_i,\delta p_i)$, which will be as usual first order in time. 
  In this particular case, one may also easily eliminate $\delta p_i$ and obtain a {\em second order} differential equation for the $\delta q$ (the `Geodesic Deviation') exclusively.
  We need here only to remark that it is of the form \cite{Hartle} $ \delta \ddot q_i = G(R, \dot q, \delta q_i)$, with $G$ quadratic in $\dot q_i$ and linear in both the Riemann curvature tensor
  $R$ and $\delta q_i$.  
 In a chaotic system, the purely spatial separation is expected to scale exponentially with the geodesic length $\ell_G$ along the trajectory (Fig. \ref{remsketch})
 \begin{equation}
 \Delta(\ell_G) \sim \Delta(0) e^{\ell_G/\ell_N}
 \end{equation}
 where $\ell_N$, the  {\em Lyapunov length}, is the typical distance travelled for the separation of two nearby trajectories to be multiplied by $e$.
 We expect that in a system with a good thermodynamic limit,  $\ell_N$ scales with the size as 
 \begin{equation}
 \ell_N = {\ell_o}{\sqrt N} 
 \end{equation}
 with $\ell_o$ a length that has a finite thermodynamic limit. We may understand this scaling as follows: if we consider the Cartesian product of two  independent  chaotic systems with the same
 Lyapunov length, and consider pairs of trajectories for both systems starting from separations $\Delta_0$.  After a length $\ell_o$ travelled by {\em each}, their separations will achieve $\Delta(t)$ for both.
 But this happens in a combined length travelled that is now $\sqrt 2 \ell_o$, while the total separation has increased only by $e$.
 
 Consider now the usual Lyapunov exponent associated with time.
 It is clear that the same pair of nearby geodesics traversed at twice the speed will diverge twice as fast.   Because the energy is quadratic in
  the speed, we conclude then that:
  \begin{equation}
  \lambda=\frac{\sqrt{2E/m}}{\ell_N}\;\;\; \Rightarrow  \;\;\; \lambda=\frac{\sqrt{{2\cal{E}}/m}}{\ell_o} \
  \label{ellzero}
  \end{equation}
  The right hand side confirms that $\ell_o$ has a good thermodynamic limit if $\lambda$ also does.  
(The largest Lyapunov value is expected to be $O(1)$ in $N$, although there have been some debate about how general this is, see \cite{Ruffo}.)
%
  A quantum system will have a thermal de Broglie length defined as \footnote{Note that a factor is not the conventional one often used.}:
 \begin{equation}
   \ell= \left(\frac{4 \pi^2 \hbar^2 }{Tm}\right)^{1/2} 
  \end{equation}
This is a quantity  {\em per degree of freedom}. The wavelength along a trajectory in full phase-space  will be $\sqrt N$ times that, i.e.   $\propto \sqrt N \ell$.
Let us compare   Lyapunov and deBroglie lengths:
  \begin{eqnarray}
 \ell_N \;\; &\div& \;\; \sqrt N \ell \nonumber \\
 \ell_0 \;\; &\div& \;\;  \ell= \left(\frac{4 \pi^2 \hbar^2 }{Tm}\right)^{1/2} \nonumber \\
  \frac{\sqrt{2E/mN}}{\lambda} \;\; &\div& \;\;  \left(\frac{4 \pi^2\hbar^2 }{Tm}\right)^{1/2} \nonumber \\
  \frac{\sqrt{T/m}}{\lambda} \;\; &\div& \;\;  \left(\frac{4 \pi^2\hbar^2 }{Tm}\right)^{1/2}
  \end{eqnarray}
  Rearranging, we find that both lengths are comparable if:
  \begin{eqnarray}
  \frac{{T}} \hbar\;\; &\sim& \;\;    {\lambda}
  \end{eqnarray}
  where we omitted the numerical factors in the last, dimensional comparison.

 The bound  (\ref{bound}) has hence  the meaning that when the thermal deBroglie 
  length is  of the order or larger than  the characteristic `chaos' length $l_o$, the semiclassical estimate breaks down and chaoticity is bounded, see Figure \ref{remsketch}.
  It is hard to scatter a wavefront along a distance shorter than its wavelength.

\section{Calculation}

Let us make this argument more precise. 
In order to do this, we make a semiclassical calculation in a way that closely resembles the one of Jalabert and Pastawski \cite{Pastawski,Prosen,Vanicek}.
We sketch the calculation here and leave the details for the Appendix.
 We have to compute  (\ref{lyap2}), the first term of which is 
\begin{equation}
T= \mbox{Tr}\left\{
 e^{i\frac{t}{\hbar}H}   B  e^{-i\frac{t}{\hbar}H} 
   e^{-\frac{\beta}{4} H } W e^{-\frac{\beta}{4} H }
   e^{i\frac{t}{\hbar}H}   B  e^{-i\frac{t}{\hbar}H} 
   e^{-\frac{\beta}{4} H } W  e^{-\frac{\beta}{4} H }  \right\}
\end{equation}

We rescale  quantities using:
\begin{equation}
t_o=\beta \hbar \hspace{1cm} ; \hspace{1cm} \ell= \left(\frac{4 \pi^2 \hbar^2 \beta}{m}\right)^{1/2}  \hspace{1cm} \hat t \equiv \frac{ t\ell}{4 \pi^2 t_o } = \frac{t}{2 \pi} \sqrt{\frac{T}{  m }} \sim {\mbox{  distance}}/2\pi\;\;\;
\end{equation}
 The value of $2 \pi \hat t$ is the typical distance
traveled by a classical particle at temperature $T$, it is much longer than the thermal length. To recap,  we have three quantities, $\ell_o$, $\ell$ and $\hat t$:
`spatial divergence', thermal deBroglie and trajectory lengths, respectively.

\begin{figure}
\centering \includegraphics[angle=0,width=5cm]{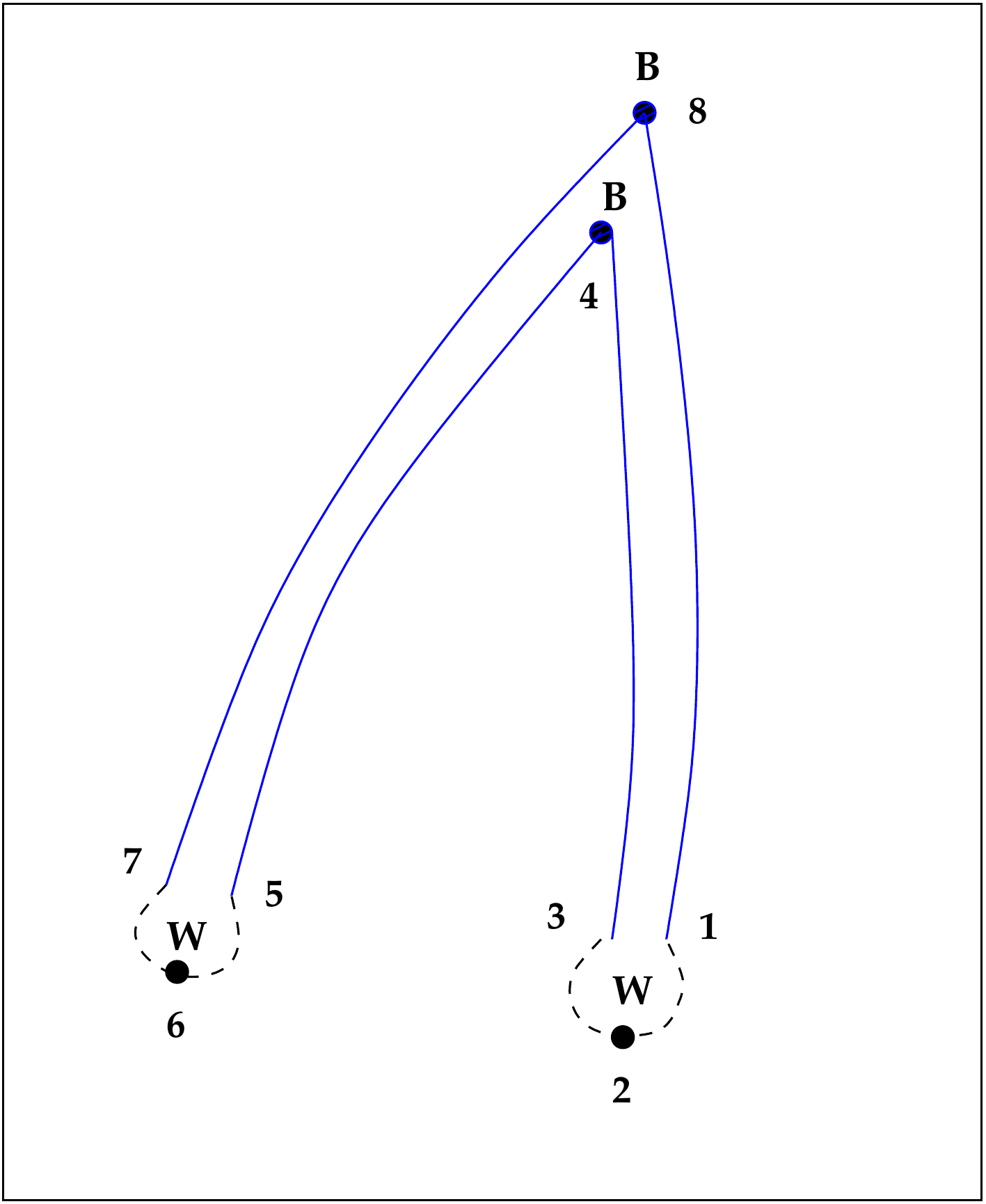} \protect\protect\protect\protect\caption{The trajectories: full and dashed lines are real-time propagation and diffusion, respectively.}
\label{trajectory} 
\end{figure}

We shall now write the real and imaginary  time propagations as:
\begin{equation}
\mbox{real time} =e^{-i\frac t\hbar H}= e^{ -\frac{i\hat t}{\ell}  [-\frac{\ell^2}{2} g^{-\frac12} \partial_i g^{ij} g^{\frac12} \partial_j]} = e^{-\frac{i\hat t}{ \ell} \bar H} \hspace{.7cm} ; \hspace{.7cm} 
\mbox{imag. time }=e^{-{\frac \beta 4} H}=e^{+ \frac{\ell^2}{32\pi^2} g^{-\frac12} \partial_i g^{ij} g^{\frac12} \partial_j } = e^{-D}
\end{equation}
which implicitly define $\bar H$ and $D$.
We recognize the real time propagations as semiclassical, with $\ell$ playing the role usually played by $\hbar$. As to the imaginary propagations,
it is diffusion in curved space with `noise' of small amplitude $\propto \ell$.
Let us introduce a basis:
\begin{eqnarray}
T= \int dq_1...dq_8 W(q_2) B(q_4)W(q_6)B(q_8)  \hspace{9cm}\nonumber\\
 \langle q_1| e^{i\frac{\hat t}{\ell}{\bar H}}|q_8\rangle     \langle q_8|e^{-i\frac{\hat t}{\ell}{\bar H}} |q_7\rangle
   \langle q_7| e^{-D } |q_6\rangle \langle q_6|e^{- D }|q_5\rangle
   \langle q_5| e^{i\frac{\hat t}{\ell}{\bar H}} |q_4\rangle     \langle q_4|e^{-i\frac{\hat t}{\ell}{\bar H}} |q_3\rangle
   \langle q_3| e^{- D }|q_2\rangle           \langle q_2|e^{- D }|q_1\rangle
\end{eqnarray}
We now use the fact that the real time propagations are semiclassical, and that the diffusions are over very small $O(\ell)$ lengths (Fig. \ref{trajectory}).
We first write the short diffusion episodes introducing $O(1)$ lengths $y_1,y_2$:
\begin{eqnarray}
 R(q_5,y_1) \equiv   \left \langle q_5+\ell \frac{y_1}{2}\left | e^{- D }  W e^{- D }\right |q_5-\ell \frac{y_1}{2}\right \rangle \;\;\;\;\;  {\mbox{and}} \;\;\;\;\;
 R(q_1,y_2)   =\left \langle q_1 + \ell \frac{y_2}{2}\left | e^{- D }    W e^{- D }\right |q_1- \ell \frac{y_2}{2}\right \rangle \; ,
\end{eqnarray}
and compute  (Appendix), to leading order in $\ell$:
\begin{eqnarray}
R(q_5,y_1) &\propto&   e^{-8 \pi^2 L(q_5, y_1)}  \;\;
\label{episode}
\end{eqnarray}
and similarly for $R(q_1,y_2)$ (one may omit constant  prefactors, as they appear in the denominator too). Note that the $y_i$ play the role of velocities in the Lagrangian $L$.
For the integral, we have:
\begin{eqnarray}
&\int& dq_1  dq_4 dq_5  dq_8 \; dy_1 dy_2  B(q_4)B(q_8)   R(q_5,y_1) R(q_1,y_2) \hspace{9cm}\nonumber\\
 & &\left \langle q_1- \ell \frac{y_2}{2}\left | e^{i\frac{\hat t}{ \ell}{\bar H}}\right |q_8\right \rangle     \left \langle q_8\left |e^{-i\frac{\hat t}{ \ell}{\bar H}} \right |q_5+\ell \frac{y_1}{2}\right \rangle
   \left \langle q_5-\ell \frac{y_1}{2}\left | e^{i\frac{\hat t}{ \ell}{\bar H}} \right |q_4\right \rangle     \left \langle q_4\left |e^{-i\frac{\hat t}{ \ell}{\bar H}} \right | q_1+\ell \frac{y_2}{2}\right \rangle 
\end{eqnarray}

Next, we introduce the Van Vleck formula for the semiclassical propagator:
\begin{equation}
\langle q'|e^{\pm i\frac{\hat t}{ \ell}{\bar H}} |q\rangle   \sim A^\pm (q',q) e^{\pm  \frac{i}{\ell}S^c(q',q)}
\end{equation}
We have to take into account the shifts of $y_1$ and $y_2$ in the brackets.  Using the fact that the derivative of the classical action $S^c$ with respect to the coordinate gives the momentum, 
 we may put:
 \begin{equation}
 S^c(q'+\ell z,q)\sim S^c(q',q)+\ell zp' \;\;\; ; \;\;\; S^c(q',q+\ell z) \sim S^c(q',q)+\ell z p 
 \end{equation}
where $p,p'$ denote the classical momentum at the time corresponding to $q$ and $q'$, respectively. 
The shifts $y_1,y_2$ in the $A^\pm$ are negligible, since they are of $O(\ell)$.   
 We then have:
\begin{eqnarray}
&\int& dq_1  dq_4 dq_5  dq_8 \; dy_1 dy_2  B(q_4)B(q_8)  \;\; \;\; e^{i p_5 y_1+ ip_1 y_2-8 \pi^2 L(q_5,y_1)-8 \pi^2 L(q_1,y_2)}\hspace{7cm}\nonumber\\
  & & A^+\left (q_1,q_8 \right) e^{\frac{i}{\ell}S^c(q_1,q_8)}    
     A^-\left(q_8, q_5\right) e^{- \frac{i}{\ell}S^c(q_8,q_5)}   
     A^+\left(q_5,q_4\right) e^{ \frac{i}{\ell}S^c(q_5,q_4)}    
      A^-\left(q_4,q_1\right) e^{- \frac{i}{\ell}S^c(q_4,q_1)}    
\end{eqnarray}

We next calculate the integrals over 
$q_4,q_5,q_8$ by saddle point evaluation.  Derivatives of the action $S^c$ with respect to $q_4,q_5, q_8$ give the momenta at the end and at the beginning of the subsequent classical path, in such a way that the classical trajectory `bounces back' three times, retracing its steps, see Fig. \ref{trajectory}.  
{\em Note that the saddle point does not fix the endpoint of the trajectory (or, equivalently, the momentum), which is  a zero-mode at the classical level.} On the other hand, we are not yet integrating over the initial position.
Denoting $q^c_4=q_F+ \ell^{\frac12}  \hat q_4$,  $q^c_5=q_I+ \ell^{\frac12}  \hat q_5$, 
and  $q_8=q_F+ \ell^{\frac12}  \hat q_8$, we compute the Gaussian  integral around the saddle point, taking care of zero-modes and of Jacobians associated with changes of  variables. After a few steps (see the Appendix) , we get

\begin{eqnarray}
T \propto &\int& dp_I  dq_I  d\hat q_4 d \hat q_5    \;\; B(q_F+\ell^{ \frac 12} \hat q_4)B(q_F) W(q_I+\ell^{ \frac 12}\hat q_5) W(q_I) \;\; e^{-\frac{1}{16 \pi^2}H(p_I,q_I)} \left|\det \Lambda^{-1}\right| \;\; \exp \left\{i{\hat t}\left[\Lambda^{-1}_{ab} \hat q_{4a} \hat q_{5b}  \right] \right\}  \label{exp0}
\end{eqnarray}
where we have defined: \begin{equation} 
\;\;\;   \frac{\partial^2 S^c}{\partial q_{Ii}\partial q_{Fj}}=\frac{\partial p_{Ii}}{\partial {q_{Fj}}} =\frac{\partial p_{Fi}}{\partial {q_{Ij}}} =  [\Lambda^{-1}]_{ij} (q_I,q_F, {\hat t}) 
\label{Lambda}
\end{equation}

The leading term is cancelled by the second term in (\ref{lyap2}). The expectations with two $\hat q$ contribute only to connected two-time correlations
that have decayed in the time regime that we are studying because chaos leads to cancellations  (see the Appendix). 
We are left with the expectation containing four $\hat q$, which, after rearrangement, may be expressed in terms of classical Poisson brackets.
Assembling all the pieces together, we get the final, purely classical, result:
\begin{eqnarray}
F_1 \propto \ell  \frac{\int dp_I   \;\; dq_I  \;\; \{B_{F},W_I\}^2(q_I, p_I, 2\pi \hat t)\;\;\; e^{-H_o(p_I,q_I)}}{ \int dp   \;\; dq_I  \;\;  e^{-H_o(p_I,q_I)}}  \label{exp1}
\end{eqnarray}
where we have defined the classical  adimensional Hamiltonian  
\begin{equation}
H_o= \frac 12 p_i p_j g^{ij} \label{adim}
\end{equation}
The meaning of this formula is clear. We are summing over trajectories of different (rescaled) speeds, starting at all points, and weighted with the  Gibbs measure.
The time is the rescaled time $\hat t$, which accounts for the absence of the temperature in the exponent, so that this expression is, for given $\hat t$,
purely geometrical. 
This is as far as we can go for a system with few degrees of freedom in the canonical framework.

\vspace{.5cm}

{\bf Canonical vs. micro-canonical}

\vspace{.5cm}

For systems with few degrees of freedom, quantities have thermal fluctuations, and  there is no single velocity associated with all trajectories in (\ref{exp1}).
Let us consider  for simplicity the case in which  the measure is concentrated on a single energy, as in the thermodynamic limit.
Introducing a delta function and rescaling with the energy density $  {\cal{E}}_o$   $p \rightarrow \sqrt{ 2 {\cal{E}}_o} p$ we have:

\begin{eqnarray}
&\int& dp   \;\; dq  \;\; \{B_{F},W_I\}^2(q,p)\;\;\; e^{-N  H_o(p,q, 4 \pi \hat t)}= \int d {\cal{E}}_o e^{-N {\cal{E}}_o} \int dq dp \; \delta( H_o(q,p) -{\cal{E}}_o)  
\;\; \{B_{F},W_I\}^2(q,p)\;\;\; \nonumber \\
&\int&  d {\cal{E}}_o e^{-N {\cal{E}}_o+\frac{N-1}2 \ln  {2 \cal{E}}_o} \int dq dp \; \delta \left( H_o(q,p)-\frac 12\right)  \;\; \{B_{F},W_I\}^2(q,\sqrt { {2 \cal{E}}_o} p, 2 \pi \hat t)\;\;\;
\label{exp2}
\end{eqnarray}
(we have dropped the subindex $`I'$ in the arguments $q,p$). Putting all together we may write:
\begin{eqnarray}
F_1 
 &\propto& \ell  \frac{\int d {\cal{E}}_o e^{-N {\cal{E}}_o+\frac{N-1}2 \ln [ 2{\cal{E}}_o]+N{\cal{S}}}  \;\;  {\bf \langle} \hspace{-.07cm}{\bf \langle} \{B_{F},W_I\}^2{\bf \rangle}\hspace{-.07cm} {\bf \rangle}(2 \pi \hat t)}{ 
 \int d {\cal{E}}_o e^{-N {\cal{E}}_o+\frac{N-1}2 \ln [2 {\cal{E}}_o]+N{\cal{S}}} }
  \end{eqnarray}
  where 
\begin{equation}
{\cal{S}} =\ln \frac{1}{N}  \int dp dq \; \delta( {\cal{E}}_o(q,p) -1/2) = \frac{1}{N}  \ln \int dp dq \; g^{\frac 12} + const
\end{equation}
is a purely geometric contribution to the entropy,
and
\begin{equation}
{\bf \langle}\hspace{-.07cm} {\bf \langle}  \{B_{F},W_I(q)\}^2 {\bf \rangle}\hspace{-.07cm} {\bf \rangle}(2 \pi \hat t) \equiv 
 \frac{\int dq dp \;  \{B_{F}(q,\sqrt{{2{\cal E}_o}}p,2 \pi \hat t),W_I(q)\}^2  \;\;\delta \left( H_o(q,p)-\frac 12\right) }{\int dq dp \;   \;\;\delta \left( H_o(q,p)-\frac 12\right) } 
 \label{vas}
 \end{equation}
The delta function imposes unit speed.
It is now clear that if the number of degrees of degrees of freedom is large, and the system has a thermodynamic limit,
a single energy density $ {\cal{E}}_o=\frac12$  will dominate and the argument of the Poisson bracket in (\ref{vas}) will simply be $(q,p)$. 
Here the limit of large $N$ has to be made {\em before} the limit of large $t$, otherwise the Poisson bracket term 
will distort the thermodynamic measure. 
For this value, we define the reference Lyapunov exponent $\lambda_o$, a purely geometric quantity
\begin{equation}
{\bf \langle}  \hspace{-.07cm}{\bf \langle}  \{B_{F},W_I\}^2 {\bf \rangle}  \hspace{-.07cm}{\bf \rangle} (2 \pi \hat t)\sim e^{\lambda_o \hat t} 
 \sim e^{  t \lambda_o \sqrt{\frac{T }{ m}} } 
 \end{equation}
which implies that $\lambda_o=\frac 1 {\ell_o}$, the spatial geodesic separation we defined above, and 
\begin{equation}
\lambda = \frac 1 {\ell_o} \sqrt{\frac T m}
\label{lyly}
\end{equation}
which is the classical estimate we made at the beginning, i.e.  the expected result that semiclassical and classical Lyapunov exponents coincide.

\begin{figure}
\centering \includegraphics[angle=0,width=5cm]{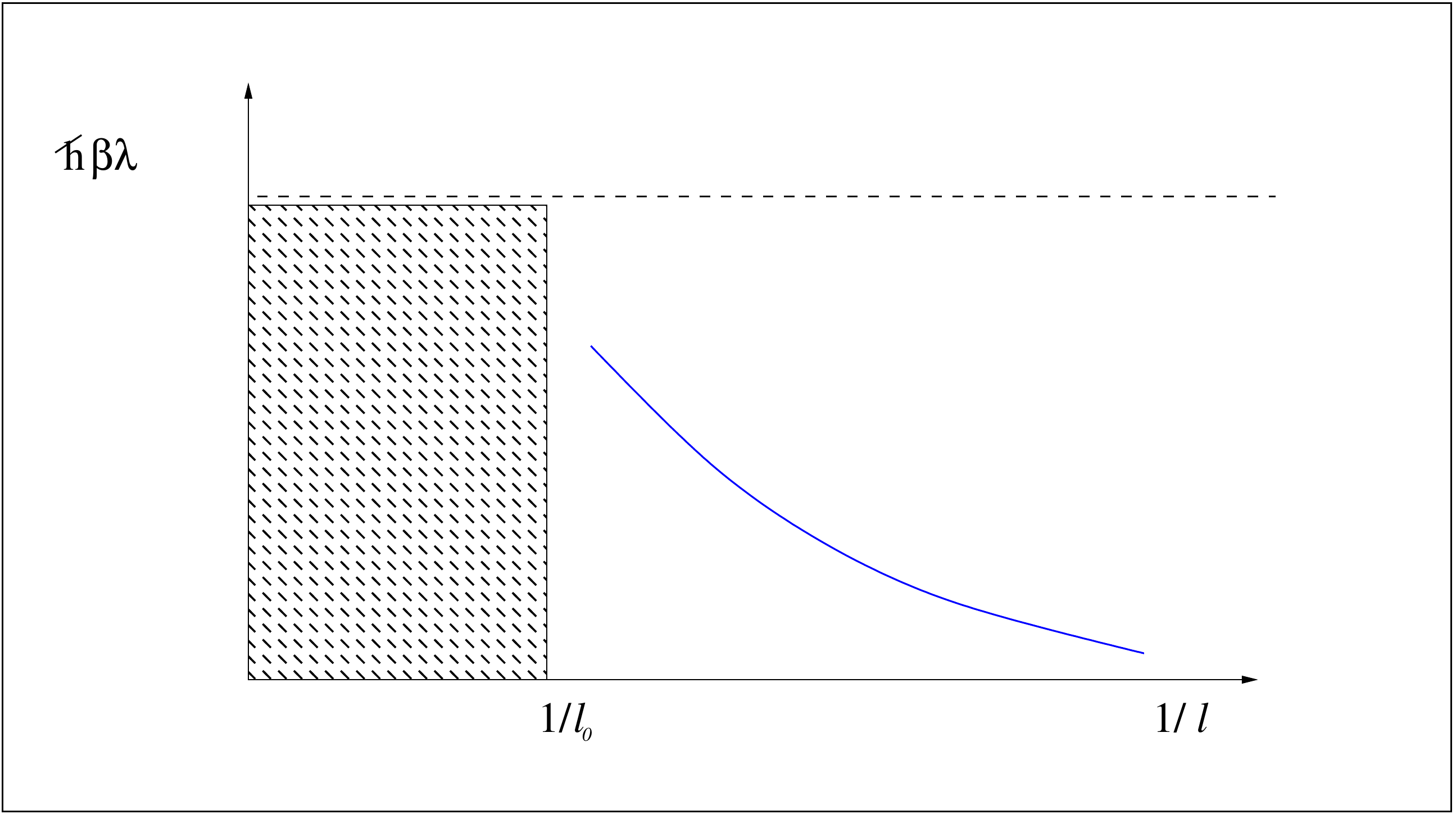} \protect\protect\protect\protect\caption{The bound acts at the point where the thermal wavelength become
of the same order as  the Lyapunov length. At longer thermal wavelengths (the shaded region) quantum dynamics dominates of \cite{Pastawski}}.
\label{trajectory1} 
\end{figure}

\begin{equation}
\end{equation}

\pagebreak

\vspace{.3cm}

{\bf Quantum length $\ell$ and  characteristic  length(s) of the geometry}

\vspace{.2cm}

The situation we have is sketched in Fig \ref{trajectory1}. For short de Broglie wavelengths, the adimensional Lyapunov exponent $\beta \hbar \lambda$ is proportional to 
$\frac{\ell}{\ell_o}$.   
If the system has a single characteristic length $R$, as for example the case of the pseudosphere, whose only parameter is its `radius' $R$,
then the classical Lyapunov exponent scales with this length, and there is only the question of how does the thermal deBroglie length compare
with the geometric one. As we consider longer quantum wavelengths, the semiclassical approximation becomes less good, and
in the regime where the wavelength becomes large with respect to all the geometrical features  completely  breaks down.
It is not even clear that a Lyapunov regime --  exponentially dependent on $t$ and  independent of $B$ and $W$ -- exists for the Loschmidt echo, and one would expect that
chaoticity becomes smaller.

In the  example of the motion on the pseudosphere, the  Hamiltonian has a continuous spectrum 
starting from a lowest eigenvalue \cite{Voros}
\begin{equation}
E_0= \frac{\hbar^2}{m \ell_o^2}=\frac{\hbar^2}{m R^2}
\end{equation}
The same system bounded to a large region, will have a similar spectrum (this time not continuous), plus a `flat' state of eigenvalue zero $|0\rangle$.
Let us compare the thermal energy with the lowest nonzero eigenvector:
\begin{equation}
\beta E_0= \frac{\hbar^2}{m T\ell_o^2}=\left(\frac{\ell}{\ell_o}\right)^2
\end{equation}
We see what happens when $\ell$ becomes larger than $\ell_o$: the factor in the exponent of $e^{-\beta H}$ projects onto the lowest, non-propagating mode.
We may think of the evolution as diffusion over a long  time $\beta$, larger than the ergodic time of the diffusive system.
We may estimate that the result of $F_1$  for $\beta E_o\gg 0$ is, up to exponentially small $(\ell/\ell_o)^2$:
\begin{equation}
F_1 =\langle B^2\rangle _{0} \langle W^2 \rangle_{0} -  \langle B\rangle_0^2 \langle W \rangle_0^2
\end{equation}
where $\langle \bullet \rangle_0$ denotes flat averages over the manifold.
Again, it seems doubtful that there is chaos in this regime $\ell \ll \ell_o$  (the gray region), and it is not even clear whether  a Lyapunov exponent independent of $W,B$, governing an exponential relaxation, still exists.

 Is it possible to have chaos all the way down to zero temperature? It is easy to construct a (very artificial) model that does. For this we have to use the fact, already mentioned above,
 that the Lyapunov exponent of the Cartesian product of two weakly or non-interacting systems is just the largest of the exponents of the individual systems.
 Consider then a $2N$ dimensional system composed of $N$  hyperbolic systems like (\ref{hyper}), each with typical
 `radius' $R_1<R_2<...<R_N$. The total Lyapunov of the combined system will be composed of the envelope of the highest  of the curves of Fig.  \ref{trajectory1}, each one 
 with a characteristic $\ell_{o1} <\ell_{o2} <...<\ell_{oN}$.  In such a case, one may imagine that as one moves into
the deeper quantum regime (longer thermal lengths), the wave propagation ignores the shorter geometric  lengths but is scattered by the longest. In a system with regions with
all levels of curvature, the quantum Lyapunov would then be governed by the  smaller curvature of the order of the inverse deBroglie length, and this is compatible
with $\beta \hbar \lambda$ stabilizing as the thermal length becomes longer and longer.
  The example of critical opalescence 
comes to mind: at exactly the critical point there are fluctuations of density -- and hence of refractive index -- at all scales, and the fluid looks white because it scatters
all wavelengths.  It is tempting to compare the situation with  systems like SYK that saturate the bound at exactly zero temperature and have a critical, gapless point there.

\section{Conclusions}
\label{Conclusion}

The example of a particle moving on a curved surface, or of a `bosonic' system with a classical ground-state manifold of high dimension, is a context where
one may picture the action of the quantum bounds on chaos in an intuitive way. It would be interesting  to discuss the effect of `driving' such a system,
and computing transport and dissipation coefficients, and the limits quantum mechanics places upon them. 

Another interesting exercise would to obtain a bound that applies to the Loschmidt echo in the more conventional setting of a wave packet with typical energy $E$, rather than in  the canonical ensemble. It seems clear that some such bound should exist, and would be useful for systems with  few degrees of freedom. 

\vspace{1.7cm}

\vspace{1cm}

{\bf \large Appendix}

\vspace{.3cm}

Let us first compute the diffusion intervals.

\begin{eqnarray}
R(q_5,y_1) &=&   \langle q_5+\ell \frac{y_1}{2}| e^{- D }  W e^{- D }|q_5-\ell \frac{y_1}{2}\rangle  \nonumber \\
&\propto&   \int dq'  e^{- \left({\frac{8 \pi^2}{\ell^2}}\right)
  g_{ij} (q_5+\ell \frac{y_1}{2}-q')_i (q_5+\ell \frac{y_1}{2}-q')_j }  W(q')  e^{-\left({\frac{8 \pi^2}{\ell^2}}\right) g_{ij}^{-1} (q_5-\ell \frac{y_1}{2}-q')_i (q_5-\ell \frac{y_1}{2}-q')_j } 
\nonumber\\
& &   \int dq'  e^{-  \left({\frac{16 \pi^2}{\ell^2}}\right)  g_{ij} (q_5-q')_i (q_5-q')_j }  \left[W(q_5)+\frac{\partial W}{\partial q_a}(q'-q_5)_a+...\right]    e^{- 4 \pi^2 g_{ij} {y_{1i} y_{1j}} }\nonumber\\
&\sim& W(q_5) e^{-{{4 \pi^2}}g_{ij}(q_5)y_{1i} y_{1j} } = e^{-{{8 \pi^2}} L_o(q_5, y_1)/2}  \;\; 
\end{eqnarray}
and similarly for $R(q_1,y_2)$. Note that the $y_i$ play the role of velocities in the adimensionalized  Lagrangian $L_o$.

 For the full expression we then have, using the semiclassical propagators:
\begin{eqnarray}
T&=& \int dq_1  dq_4 dq_5  dq_8 \; dy_1 dy_2  B(q_4)B(q_8)  \;\;\;\;  \;\; e^{i p_5 y_1+ ip_1 y_2 -  8 \pi^2 L_o(q_5,y_1)- 8 \pi^2 L_o(q_1,y_2)} \; A^+\left (q_1-\ell \frac{y_2}2,q_8 \right) e^{\frac{i}{\ell}S^c(q_1,q_8)}  \nonumber\\
& &     
     A^-\left(q_8, q_5+\ell \frac{y_1}2\right) e^{- \frac{i}{\ell}S^c(q_8,q_5)}   
     A^+\left(q_5-\ell \frac{y_1}2,q_4\right) e^{ \frac{i}{\ell}S^c(q_5,q_4)}    
      A^-\left(q_4,q_1+\ell \frac{y_2}2\right) e^{- \frac{i}{\ell}S^c(q_4,q_1)}    
\end{eqnarray}

We next calculate the integrals over 
$q_4,q_5,q_8$ by saddle point evaluation.  We get:
\begin{equation}
q^c_1=q^c_5=q_I  \;\;\; and \;\;\; q^c_8=q^c_4=q_F
\end{equation}
As mentioned above, derivatives of the action $S^c$ with respect to $q_4,q_5, q_8$ give the momenta at the end and at the beginning of the subsequent classical path, in such a way that the classical trajectory `bounces back' three times, retracing its steps see Fig. \ref{trajectory}. The speed is the same in modulus
in the four trajectories, but is as yet undetermined.  Or, in other words, 
the saddle point does not fix the endpoint of the trajectory, which is then  a zero-mode at the classical level.

Denoting $q^c_4=q_F+ \ell^{\frac12}  \hat q_4$,  $q^c_5=q_I+ \ell^{\frac12}  \hat q_5$, 
and  $q_8=q_F+ \ell^{\frac12}  \hat q_8$, we compute the integral around the saddle point.  Rather than expressing it as a path integral, we write an expression for 
the Gaussian fluctuations around each trajectory. 
For this we expand 
\begin{eqnarray}
S^c(q_I+\ell^{\frac 12} \hat q',q_F+ \ell^{\frac 12} \hat q)&=& S^c(q_l,q_F) + {\mbox{...linear...}}  + \frac{\ell}{2}  \left[\frac{\partial^2 S^c}{\partial q^I_a \partial q^I_b}\hat q_a \hat q_b +
\frac{\partial^2 S^c}{\partial q^I_a \partial q^F_b}\hat q_a \hat q_b'+\frac{\partial^2 S^c}{\partial q^F_a \partial q^F_b}\hat q_a' \hat q_b'\right] \nonumber \\
   A^+\left( q_1-\ell \frac{y_2}2,q_8 \right) &\sim& A^+(q_I,q_F)=A^-(q_F,q_I)
   \end{eqnarray}
where, as usual,  linear terms will cancel at the saddle point, and second derivatives are evaluated on the saddle-point values $q_I,q_F$. We have kept the leading
order in the $A^\pm$.
 Note that
\begin{equation}
\det^{-1}[\Lambda] = |A(q_I,q_F)|^2
\end{equation}


We now develop the classical action up to second order in $\hat q_4, \hat q_8$, and $\hat q_5$ in the exponent. We also need to integrate  over  the multiple saddle points, each with a given $q_F$. The Gaussian fluctuations naturally contain zero-modes associated to
the multiplicity of saddles, we must supress them   with a factor $\delta(\hat q_8)$, otherwise we would be double-counting the integral over $q_F$.  All in all, we get:
\begin{eqnarray}
&\int& dq_F   \;\; dq_I  d\hat q_4 d \hat q_5   \; dy_1 dy_2  \;\; B(q_F+\ell^{ \frac 12} \hat q_4)B(q_F) W(q_I+\ell^{ \frac 12}\hat q_5) W(q_I) \;\; e^{i p_5 y_1+ ip_1 y_2-8 \pi^2L_o(q_I,y_1)-
8 \pi^2 L_o(q_I,y_2)}\hspace{7cm}\nonumber\\
& &   \left|A(q_I,q_F)\right|^4 \;\; \exp \left\{i{\hat t}\left[\Lambda_{ab} \hat q_{4a} \hat q_{5b}  \right] \right\}  
     \end{eqnarray}Note the cancellations of terms with second derivatives with respect to a single vector $\frac{\partial^2 S^c}{\partial  \hat q_{5i} \partial \hat q_{5j}}$, etc., and also
     the absence of terms with $\hat q_8$ due to the $\delta(\hat q_8)$.
     We now integrate over $y_1$ and $y_2$, which transforms $L_o$ into the adimensional Hamiltonian (\ref{adim}) in
the exponent,  to obtain:
\begin{eqnarray}
&\int& dq_F   \;\; dq_I  d\hat q_4 d \hat q_5    \;\; B(q_F+\ell^{ \frac 12} \hat q_4)B(q_F) W(q_I+\ell^{ \frac 12}\hat q_5) W(q_I) \;\; e^{-\frac{1}{4 \pi^2} H_o(p_I,q_I)}
  \left|A(q_I,q_F)\right|^4 \;\; \exp \left\{i{\hat t}\left[\Lambda^{-1}_{ab} \hat q_{4a} \hat q_{5b}  \right] \right\}  
\end{eqnarray}
The final step is to change integration variables from $(q_I,q_F)$ to $(q_I,p)$:
\begin{equation}
dp = \left|A(q_I,q_F)\right|^2 dq_F
\end{equation}
so that:
\begin{eqnarray}
T= &\int& dp   \;\; dq_I  d\hat q_4 d \hat q_5    \;\; B(q_F+\ell^{ \frac 12} \hat q_4)B(q_F) W(q_I+\ell^{ \frac 12}\hat q_5) W(q_I) \;\; e^{-\frac{1}{4\pi^2}H_o (p_I,q_I)}    \det[\Lambda^{-1}  ]                \;\; \exp \left\{i{\hat t}\left[\Lambda^{-1}_{ab} \hat q_{4a} \hat q_{5b}  \right] \right\}  \label{exp}
\end{eqnarray}     
Note that the normalization is a constant independent of $\Lambda$.

Let us expand
\begin{equation}
W(q_c+\ell^{\frac 12} \hat q_5) = W(q_I)+ \ell^{\frac12}\frac{\partial W}{\partial q_j}^c \hat q_{5j}+ \ell  \frac{\partial^2 W}{\partial q_j \partial q_k} \hat q_{5j} \hat q_{5k} +...    \;\;\;\; ; \;\;\; 
 B(q_F+\ell^{\frac 12} \hat q_{4j}) = B(q_c) + \ell^{\frac12}\frac{\partial B}{\partial q_j}^c  \hat q_{4j} +\ell  \frac{\partial^2 B}{\partial q_j \partial q_k} \hat q_{4j} \hat q_{4k} +...
\end{equation}

Substituting  these developments  in (\ref{exp}) there are several terms that contribute:

$\bullet$ The leading term proportional to $B^2(q_F)W^2(q_I)$, which is cancelled by the corresponding one coming from the second term in (\ref{lyap2}).

$\bullet$ Expectations of terms of the form  $B(q_F)W(q_I)W_{,i}(q_I)B_{,j}(q_F) \langle q_{5i} q_{4j} \rangle$ (comma denotes derivative) are proportional to $\Lambda_{kl}$. They are zero at long times
for classical reasons: the basis that diagonalizes $\Lambda_{ij}$ varies wildly from trajectory to trajectory in a chaotic system. This is how the two-time connected correlation function
vanishes at the times we are interested in. Similar two-time contributions  come from  the second term in (\ref{lyap2}), and are negligible for the same reason.

$\bullet$ Finally, the relevant contribution, of $O(\ell)$ comes from:
 \begin{equation}
 B(q_F)W(q_I)W_{,ij}(q_I)B_{,kl}(q_F) \langle \hat q_{5i} \hat q_{5j} \hat q_{4k} q_{4l} \rangle = - B(q_F)W(q_I)W_{,ij}(q_I)B_{,kl}(q_F) \Lambda_{ik}\Lambda_{jl}
\end{equation}
Up to a total derivative in $q_I$ and $q_F$ this yields, neglecting rapidly oscillating terms 
 \begin{equation}
 B_{,l}(q_{F})W_{,j}(q_I)W_{,i}(q_I)B_{,k}(q_F) \Lambda_{ik}\Lambda_{jl}=\{B(q_{F}),W(q_I)\}^2 \end{equation}
 where we have used
\begin{equation}
 B_{,l} \Lambda_{jl}W_{,j}= \frac{\partial B(q_F)}{\partial q_{Fl}} \frac{\partial q_{Fl}}{\partial p_{Ij}}\frac{\partial W(q_I)}{\partial q_{Ij}}=\frac{\partial B(q_F)}{\partial p_{Fj}} \frac{\partial W(q_I)}{\partial q_{Ij}} =-\{B(q_F),W(q_I)\}
 \end{equation}

Putting all the pieces together, we get
\begin{eqnarray}
F_1 \propto \ell  \frac{\int dp_I   \;\; dq_I  \;\; \{B_{F},W_I\}^2(q_I,p_I,{\hat t})\;\;\; e^{-\frac{1}{4 \pi^2}H_o(p_I,q_I)}}{ \int dp   \;\; dq_I  \;\;  e^{-\frac{1}{4 \pi^2}H_o(p_I,q_I)}}  \label{exp3}
\end{eqnarray}
 rescaling  $p \rightarrow 2 \pi p$ and then $\hat t \rightarrow 2 \pi \hat t$, we get rid of the prefactor in the exponent and we obtain the final result   (\ref{exp1}).

\vspace{1cm}

{\bf \large Aknowledgments}

\vspace{1cm}

 I would like to thank  C. Bachas,  D Bernard,  G Biroli and D. Reichman
for helpful discussions, and J Maldacena for relevant references.
This work was supported by a grant from the Simon Foundation (No 454943).


\begin{thebibliography}{100}
\bibitem{mss}Maldacena, Juan, Stephen H. Shenker, and Douglas Stanford. "A bound on chaos." arXiv preprint arXiv:1503.01409 (2015).
\bibitem{Hayden}Hayden, Patrick, and John Preskill. "Black holes as mirrors: quantum information in random subsystems." Journal of High Energy Physics 2007.09 (2007): 120.
\bibitem{Susskind}Sekino, Yasuhiro, and Leonard Susskind. "Fast scramblers." Journal of High Energy Physics 2008.10 (2008): 065.
\bibitem{Campisi}Campisi, Michele, and John Goold. "Thermodynamics of the quantum butterfly effect." arXiv preprint arXiv:1609.07848 (2016).
\bibitem{Levstein} For experimental evidence, see: Levstein, Patricia R., Gonzalo Usaj, and Horacio M. Pastawski. "Attenuation of polarization echoes in nuclear magnetic resonance: A study of the emergence of dynamical irreversibility in many-body quantum systems." The Journal of chemical physics 108.7 (1998): 2718-2724.
\bibitem{Miyaji} Miyaji, Masamichi. "Butterflies from information metric." arXiv preprint arXiv:1607.01467 (2016).
\bibitem{Peres}Peres, Asher. "Stability of quantum motion in chaotic and regular systems." Physical Review A 30.4 (1984): 1610.
\bibitem{Pastawski}Jalabert, Rodolfo A., and Horacio M. Pastawski. "environment-independent decoherence rate in classically chaotic systems." Physical review letters 86.12 (2001): 2490.
\bibitem{Prosen}See: Gorin, Thomas, et al. "Dynamics of Loschmidt echoes and fidelity decay." Physics Reports 435.2 (2006): 33-156, and references therein.
\bibitem{Vanicek}Vanicek, Jiri, and eric J. Heller. "Semiclassical evaluation of quantum fidelity." Physical Review e 68.5 (2003): 056208.
\bibitem{Beenakker} Silvestrov, P. G., J. Tworzydo, and C. W. J. Beenakker. "Hypersensitivity to perturbations of quantum-chaotic wave-packet dynamics." Physical Review e 67.2 (2003): 025204.
\bibitem{Garcia}Berenstein, David, and Antonio M. Garc\'ia-Garc\'ia. "Universal quantum constraints on the butterfly effect." arXiv preprint arXiv:1510.08870 (2015).
\bibitem{wires}Jensen, H., and H. Koppe. "Quantum mechanics with constraints." Annals of Physics 63.2 (1971): 586-591.
\bibitem{Kitaev}A. Kitaev, A simple model of quantum holography."
http://online.kitp.ucsb.edu/online/entangled15/kitaev/,http:
//online.kitp.ucsb.edu/online/entangled15/kitaev2/. Talks at KITP, April
7, 2015 and May 27, 2015.
\bibitem{ms}Maldacena, Juan, and Douglas Stanford. "Comments on the Sachdev-Ye-Kitaev model." arXiv preprint arXiv:1604.07818 (2016).

\bibitem{Voros}Balazs, N. L., and A. Voros. "Chaos on the pseudosphere." Physics reports 143.3 (1986): 109-240.
\bibitem{Ikegami}  
Let us note in passing that there is some arbitrariness in terms of order $\hbar$ that depend on how we fix the constraints, here
for simplicity we choose the Laplace Beltrami operator: see
Ikegami, Mitsuhiro, et al. "Quantum Mechanics of a Particle on a Curved Surface Comparison of Three Different Approaches." Progress of theoretical physics 88.2 (1992): 229-249.
\bibitem{syk}Sachdev, Subir, and Jinwu Ye. "Gapless spin-fluid ground state in a random quantum Heisenberg magnet." Physical review letters 70.21 (1993): 3339.\\
Parcollet, Olivier, and Antoine Georges. "Non-Fermi-liquid regime of a doped Mott insulator." Physical Review B 59.8 (1999): 5341.\\
Polchinski, Joseph, and Vladimir Rosenhaus. "The spectrum in the Sachdev-Ye-Kitaev model." arXiv preprint arXiv:1601.07768 (2016).
\bibitem{Ruffo}Livi, R., A. Politi, and S. Ruffo. "Distribution of characteristic exponents in the thermodynamic limit." Journal of Physics A: Mathematical and General 19.11 (1986): 2033.
\bibitem{Hartle} Hartle, James B. Gravity: an introduction to einstein's general relativity. Vol. 1. 2003.
\bibitem{butterfly}Shenker, Stephen H., and Douglas Stanford. "Black holes and the butterfly effect." arXiv preprint arXiv:1306.0722 (2013).
\bibitem{Miller}Miller, William H., Steven D. Schwartz, and John W. Tromp. "Quantum mechanical rate constants for bimolecular reactions." The Journal of chemical physics 79.10 (1983): 4889-4898.
\end{thebibliography}
\end{document}